\documentclass[article, sort&compress]{elsarticle}

\usepackage{hyperref}
\usepackage{graphicx}
\usepackage{amsmath}
\usepackage{amssymb}
\usepackage{color}
\usepackage[normalem]{ulem}
\usepackage{lineno}

\newcommand{\of}[1]{\left(#1\right)}

\newcommand{\mn}[1]{\langle #1\rangle}

\journal{Chaos, Solitons and Fractals}

\begin{document}

\begin{frontmatter}

\title{Effects of correlated noise on the excitation of robust breathers in an ac-driven, lossy sine-Gordon system}


\author[unipa]{Giovanni Di Fresco}

\author[unipa]{Duilio De Santis\corref{correspondingauthor}}
\cortext[correspondingauthor]{Corresponding author}
\ead{duilio.desantis@unipa.it}
\author[unisa,INFN]{Claudio Guarcello}
\author[unipa,unn]{Bernardo Spagnolo}
%
\author[unipa]{Angelo Carollo}
\author[unipa]{Davide Valenti}
\address[unipa]{Dipartimento di Fisica e Chimica ``E.~Segr\`{e}", Group of Interdisciplinary Theoretical Physics, Università degli Studi di 
Palermo, I-90128 Palermo, Italy}
\address[unisa]{Dipartimento di Fisica ``E.~R.~Caianiello", Università degli Studi di Salerno, I-84084 Fisciano, Salerno, Italy}
\address[INFN]{INFN, Sezione di Napoli, Gruppo Collegato di Salerno - Complesso Universitario di Monte
S. Angelo, I-80126 Napoli, Italy}
\address[unn]{Stochastic Multistable Systems Laboratory, Lobachevsky University, 603950 Nizhniy Novgorod, Russia}

\begin{abstract}
Thermal noise and harmonic forcing have recently been shown to cooperatively excite sine-Gordon breathers robust to dissipation. Such a phenomenon has been found assuming a Gaussian noise source, delta-correlated both in time and space. In light of the potential implications of this generation technique, e.g., for the experimental observation of breathers in long Josephson junctions, it is physically motivated to investigate the effects of more realistic noise sources with finite correlation time and/or correlation length. Here, breathers are demonstrated to still emerge under this broader class of noise sources. The correlation time and the correlation length are found to offer control over the probability of observing breathers, as well on the typical timescale for their emergence. In particular, our results show that, as compared to the thermal case, the temporal and spatial correlations in the noise can lead to a larger breather-only occurrence frequency, i.e., the latter quantity behaves nonmonotonically versus both the correlation time and the correlation length. Overall, noise correlations represent a powerful tool for controlling the excitation of the elusive breather modes in view of experiments.
\end{abstract}

\begin{keyword}
Perturbed sine-Gordon model \sep Stochastic processes \sep Noise-induced breathers \sep Correlated noise \sep Ornstein–Uhlenbeck process
\end{keyword}

\end{frontmatter}


\section{Introduction}
\label{intro}

Many remarkable phenomena can arise thanks to the combined action of nonlinearity and noise, and their understanding is crucial for advancing both theoretical and practical knowledge in various physical systems~\cite{SanMiguel_2000,Scott_2006,Garcia_2012}. While white Gaussian (i.e., thermal) noise is often used as a simplifying abstraction, real-world applications demand a more nuanced exploration of correlated noise, that is, a stochastic source characterized by an intrinsic correlation time/length stemming from the physical processes underlying the noise's description~\cite{Sancho_1989,Haunggi_1994,SanMiguel_2000,Garcia_2012,Maggi_2022,Tiokang_2022}. This is particularly relevant, e.g., in the study of Josephson systems, which have attracted significant interest due to their fundamental properties and cutting-edge applications~\cite{Barone_1982,Pedersen_1986,Parmentier_1993,Ustinov_1998,Mazo_2014,Soloviev2014, Tafuri_2019, Braginski2019, Revin2020, Davydova2022, Nadeem2023, Guarcello2024, Ghosh2024, Hovhannisyan2024, Citro2024, Benz2024, Krasnov2024}. Within the latter context, the sine-Gordon (SG) model and its soliton excitations, (anti)kinks and breathers, are well-known to play a prominent role~\cite{Parmentier_1993,Ustinov_1998,Scott_2003,Dauxois_2006,Mazo_2014, Fujii2008,Gordeeva2010,Krasnov2011,DESANTIS2022,Sundaresh23}.

The nonlinear science community has long recognized the importance of developing a deep theoretical understanding of the SG model and studying the dynamics of its solitonic modes under harmonic (ac) forcing and dissipation~\cite{Bishop_1983,Bishop_1986,Lomdahl_1986,Kivshar_1989,Lomdahl_1988,Jensen_1991,Jensen_1992,Gro92,Kivshar_1994,Quintero_1998,Zharnitsky_1998,Rasmussen_1999,Salerno_2002,Zamora_2006,Alvarez_2014,Kiselev_2021,De_Santis_2023,De_Santis_2024_STAT}, as well as other deterministic and stochastic perturbations~\cite{Pascual_1995,Jimenez_2004,De_Santis_2023_NES}. This comprehension is not only fundamental for theoretical physics, but it could also be relevant for, e.g., gaining insight into recent pioneering experiments aimed at controlling the superconducting-like properties of strongly driven cuprate materials~\cite{Dienst_2013,Laplace_2016,Fava_2024}. In this regard, highly tunable experimental platforms such as long Josephson junctions (LJJs) and ultracold atoms~\cite{Gritsev_2007,Gritsev_2007_spectroscopy,DallaTorre_2013,Neuenhahn_2012,Lovas_2022,Wybo_2023} are particularly attractive for researchers, since they allow to replicate and investigate the properties of the SG model and its nonlinear excitations in a manageable environment, thus helping in shedding light on intricate phenomena possibly occurring in, e.g., the previously mentioned high-$T_c$ scenarios~\cite{Dienst_2013,Laplace_2016,Fava_2024}.

Motivated by the above ideas, in this work we numerically investigate the influence of both temporally and spatially correlated noise on the excitation of quasi-stable oscillatory SG soliton modes, i.e., breathers, in the presence of a spatially uniform ac driving and dissipation. Indeed, it has been shown that breathers, robust to dissipation (i.e., not decaying in time due to radiative losses) and resonant with an external ac force, can stochastically emerge if a thermal noise of suitable intensity is included in the dynamics, starting from a soliton-free initial state~\cite{De_Santis_2023,De_Santis_2024_STAT}. Here we find that this interesting phenomenon still occurs under the influence of more realistic, correlated noise sources. Furthermore, the probability of observing breathers, as well as the timescale for the excitation of these nonlinear modes, are found to be sensitive to both the correlation time $\tau$ and the correlation length $\lambda$ characterizing the noise. Notably, breathers can be observed much more frequently, as compared to the corresponding white-noise case (i.e., for $\tau,\lambda\to0$), in intermediate ranges of correlation times and correlation lengths. For both large enough $\tau$ and $\lambda$, we witness an increasing slowdown of the system's dynamics, i.e., breathers form due to fluctuations over longer timescales. These facts imply that the frequency of breather-only occurrence, defined as the fraction of the total numerical realizations in which exclusively breather modes emerge, is nonmonotonic as a function of both $\tau$ and $\lambda$. Noise correlations thus represent a tool for controlling the excitation of breathers--a useful fact in view of experiments, such as those aimed at breather detection in LJJs. We also note interesting qualitative differences in the events leading to the formation of solitonic structures between the temporally and the spatially correlated scenarios. In the former case, indeed, (regardless of $\tau$) isolated localized modes are mostly generated thanks to intense enough local fluctuations of the SG field. In the spatially correlated scenario, however, (for large enough $\lambda$) a clear tendency towards a collective behavior emerges, with easy-to-spread fluctuations leading to a cascade of solitons excited throughout the system. Nevertheless, a number of isolated breathers, robust over very long times, can still often come out of such convoluted transients.

The following outline is employed for this work. Section~\ref{mat} describes the model under consideration. Results for the temporally and the spatially correlated noise are presented, respectively, in Sections~\ref{time} and~\ref{space}. Conclusions are then drawn in Section~\ref{end}. Finally, technical details on how we numerically deal with temporally and spatially correlated sources are provided in, respectively,~\ref{app-a} and~\ref{app-b}.

\section{Materials and methods}
\label{mat}

We examine the following SG-type equation, in dimensionless units,
\begin{equation}
    \varphi_{xx} - \varphi_{tt} -\alpha\varphi_t = \sin\varphi - A \sin\of{\omega t} - \gamma\of{x,t}, \label{1}
\end{equation}
which features both deterministic and stochastic perturbations: a viscous damping term with strength $\alpha<1$, an external monochromatic (ac) force with frequency (amplitude) $\omega<1$ ($A<1$), and a space-time Gaussian noise source $\gamma\of{x,t}$ with zero average and exponential correlation, see Eqs.~\eqref{2} and~\eqref{3}. The notation $\varphi_\mu = \partial_\mu \varphi$ is used throughout to denote the partial derivatives of the field $\varphi(x,t)$. Equation~\eqref{1} is numerically integrated over the spatio-temporal domain $ (x,t) \in [0, L = 50] \times [0, T]$, via implicit finite-difference numerical means~\cite{Ames_1977,Press_1992}, with null initial conditions ($\varphi|_{t=0} = \varphi_t |_{t=0} = 0 $) and with no-flux boundary conditions ($\varphi_x |_{x=0} = \varphi_x |_{x=L} = 0 $).

We note that Eq.~\eqref{1} can be experimentally studied within various platforms, e.g., in LJJs~\cite{Barone_1982,Pedersen_1986,Parmentier_1993,Ustinov_1998,Mazo_2014}, in which case it is interpreted as the equation of motion for the Josephson phase, in the presence of dissipation, an applied uniform ac current, and stochastic fluctuations~\cite{Barone_1982,Pedersen_1986,Parmentier_1993,Ustinov_1998,Mazo_2014}.  In our work, we are interested in two types of excitations which can be studied in LJJs: solitons, which are quanta of magnetic flux, and breathers, which are bound states between flux and antiflux quanta. Furthermore, for the LJJ system, the following identifications can be made. The friction coefficient $\alpha = G/\left(\omega_p C\right)$ is related to the effective normal conductance $G$, the capacitance per unit length $C$, and the Josephson plasma frequency $\omega_p = \sqrt{2\pi J_c/\left(\Phi_0 C\right)}$, whose multiplicative inverse sets the time unit. Here, $J_c$ represents the critical Josephson current, and it determines the unit for $A$. The Josephson penetration depth $\lambda_J = \sqrt{\Phi_0/\left(2\pi J_cL_p\right)}$ sets the space unit, where $L_p$ is the inductance per unit length. The null initial condition then signifies that the LJJ is initially free of solitonic excitations. The choice of boundary conditions specifies the junction geometry under consideration---in our case, the so-called overlap geometry, in the absence of external magnetic fields~\cite{Barone_1982,Pedersen_1986,Parmentier_1993,Ustinov_1998,Mazo_2014}.

The purpose of this work is to investigate the effects of both temporal and spatial correlations in $\gamma\of{x,t}$ on the noise-induced formation of stabilized breathers, a phenomenon which was discovered in the thermal (i.e., delta-correlated) case in Refs.~\cite{De_Santis_2023,De_Santis_2024_STAT}. Going beyond the thermal noise analysis is physically motivated since, in any realistic situation, intrinsic correlation times and lengths are expected to stem from the physical processes underlying the noise's description. Thus, in Sec.~\ref{time}, we report results for a temporally correlated source defined as
\begin{equation}
 \mn{\gamma\of{x,t}\gamma\of{x^\prime, t^\prime}} = \frac{\varepsilon}{\tau} \delta\of{x - x^\prime} e^{-\frac{|t-t^\prime|}{\tau}} .\label{2}
\end{equation}
Here, $\varepsilon$ is the noise amplitude and $\tau$ is the correlation time. For a $0$-dimensional system, such correlations are known to lead to persistent phenomena in the dynamical behavior~\cite{Sancho_1989,Haunggi_1994,SanMiguel_2000,Garcia_2012,Maggi_2022,Tiokang_2022}. On the other hand, for small $\tau$, Eq.~\eqref{2}'s properties resemble that of thermal noise, which is strictly obtained in the $\tau \rightarrow 0$ limit, see~\ref{app-a} for the technical details. In this case, the system dynamics approaches that described in Refs.~\cite{De_Santis_2023,De_Santis_2024_STAT}.

Since our model is $1$-dimensional, and our focus lies on solitonic modes such as breathers, it is natural to study the influence of spatial correlations on the system dynamical features as well. In the latter scenario, which is discussed in Sec.~\ref{space}, we take
\begin{equation}
 \mn{\gamma\of{x,t}\gamma\of{x^\prime, t^\prime}} = \frac{\varepsilon}{\lambda}e^{-\frac{|x-x^\prime|}{\lambda}} \delta\of{t - t^\prime}, \label{3}
\end{equation}
where $\varepsilon$ denotes the noise amplitude and $\lambda$ is the correlation length. Spatial correlations have been introduced and studied in extended systems in the past~\cite{Garcia_2012}, but to our knowledge this topic has not been explored so far in the SG context. Before closing this section, we note that as $\lambda$ approaches zero, we revert to white noise, see~\ref{app-b} for a technical discussion on Eq.~\eqref{3}. In the latter limit, results from Refs.~\cite{De_Santis_2023,De_Santis_2024_STAT} are recovered once again.

\section{Results}
\label{res}

\subsection{Effects of temporally correlated noise}
\label{time}
\begin{figure}[t!]
    \centering
    \includegraphics[width=1\linewidth]{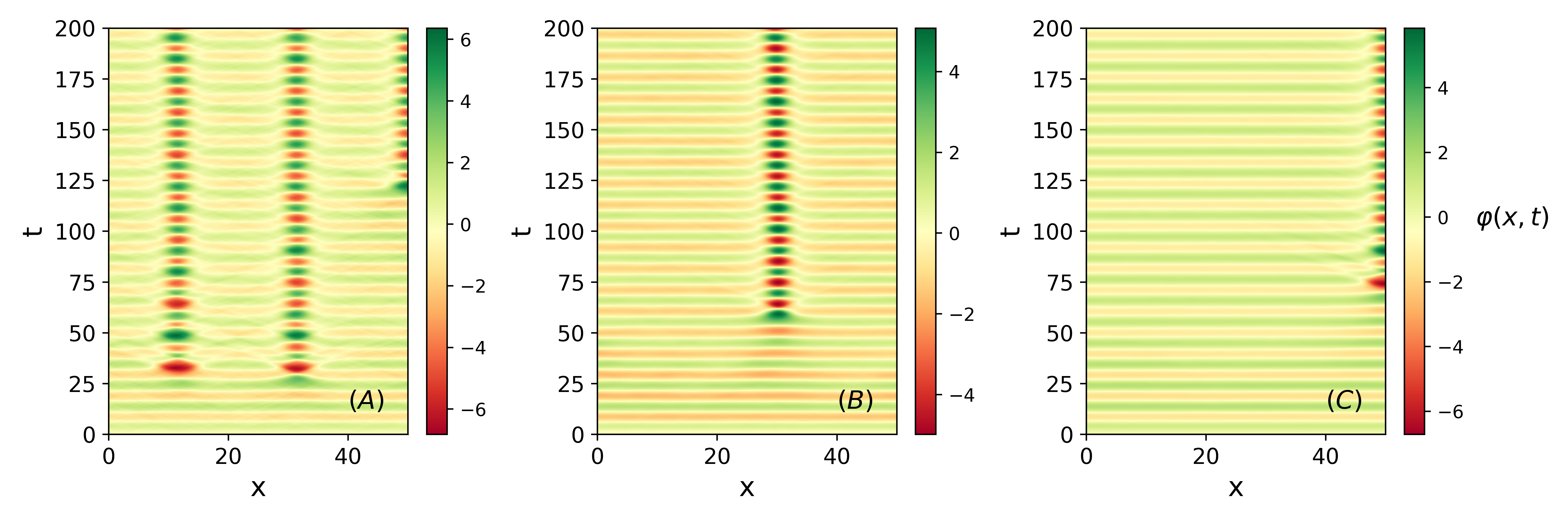}
    \caption{Contour plots of the field $\varphi\of{x, t}$, for different choices of the correlation time $\tau$, showing the robustness of the breather emergence phenomenon, as well as the $\tau$-induced slowdown of the dynamics. We take $\tau = \Delta t$ (i.e., white-noise limit) in panel $(A)$, $\tau = 300\,\Delta t$ in panel $(B)$ and $\tau = 900\Delta t$ in panel $(C)$. The remaining simulation parameters are: $\alpha = 0.2,\, A = 0.59,\, \omega = 0.6,\, \varepsilon = 0.004$.} 
    \label{fig1}
\end{figure}

In this section, the effects of temporally correlated noise, with correlation time $\tau$, are investigated~\footnote{Here, the values of $\tau$ are chosen much smaller than the simulation time, $T=500$, such that the effects of time correlations on the dynamics can be properly addressed, see~\ref{app-a}.}. To this end, we begin in Fig.~\ref{fig1} by illustrating the typical spatio-temporal behavior of the field $\varphi\of{x, t}$ for different values of $\tau$, in particular $\tau = \Delta t$ [within our discretized framework for the integration of Eq.~\eqref{1}, a $\tau$ equal to the integration time step value corresponds to the white-noise limit] is taken in panel $(A)$, $\tau = 300\,\Delta t$ in $(B)$, and $\tau = 900\,\Delta t$ in $(C)$. The remaining parameters are set such that Ref.~\cite{De_Santis_2023,De_Santis_2024_STAT}'s results are progressively reached as $\tau \to 0$, see Fig.~\ref{fig1}. A first, nontrivial fact we can deduce from these contour plots is that the emergence of breathers, remarkably stable both in amplitude and in position, is clearly possible in the presence of temporal correlations in the noise source. The kink-antikink bound state can be easily spotted in Fig.~\ref{fig1} due to its characteristic localized oscillation between $2\pi$ and $-2\pi$. Furthermore, as we progress from panel $(A)$ to panel $(C)$, i.e., as $\tau$ gets larger, the characteristic time at which the breather dynamics kicks-in seems to increase, see also Fig.~\ref{fig3}(B) below. By examining the correlation function given in Eq.~\eqref{2}, we can heuristically understand why such a phenomenon takes place as a function of the correlation time. As $\tau$ increases, the noise amplitude, which features the prefactor $\varepsilon/\tau$, is effectively reduced. Since the timescale characterizing the stochastic generation events is sensitive to the effective fluctuations' strength~\cite{De_Santis_2023,De_Santis_2024_STAT} (that is, the higher the noise intensity, the faster the coherent excitations tend to arise from the noise source), a delay in the breathers' emergence due to temporal correlations is a reasonable simulation outcome~\footnote{This phenomenon is conceptually similar to that discussed in, e.g., Ref.~\cite{Guarcello_2015_graphene}.}.

To provide statistical evidence of the phenomena exemplified in Fig.~\ref{fig1}, we perform $N=500$ independent numerical runs, and classify them according to the excitations observed in each simulation: (i)~at least a kink-type structure, if at least a $2\pi$-step excitation is present; (ii)~breathers alone, if the observed modes’ amplitudes lie between $\varphi^\star = 4\arctan\left(\sqrt{1-\omega^2}/\omega\right)$, corresponding to the amplitude of the static SG breather (with frequency $\omega$)~\cite{Scott_2003,Dauxois_2006}, and $2\pi$; (iii)~no solitonic modes, if the phase profile is essentially flat over the spatial domain. We define frequencies of occurrence for each of the previous scenarios: $f_k$, $f_b$, and $f_0$, respectively. We then analyze these occurrence frequencies under different conditions and for different ranges of parameters. Another useful tool for the present purposes is the time $t^\star$ where, in each numerical experiment, $|\varphi\of{x, t}|$ reaches (at any $x$) the threshold $\varphi^\star$. Since, for the explored parameter sets, the reaching of $t = t^\star$ reasonably signals the hitting of a solitonic generation event, here this quantity is dubbed ``hitting time'' (the convention $t^\star = \infty $ is chosen for numerical runs where $|\varphi|<\varphi^\star$ is observed throughout, like in the cases free of solitonic modes).

\begin{figure}[t!]
     \centering
     \includegraphics[width=1.0\linewidth]{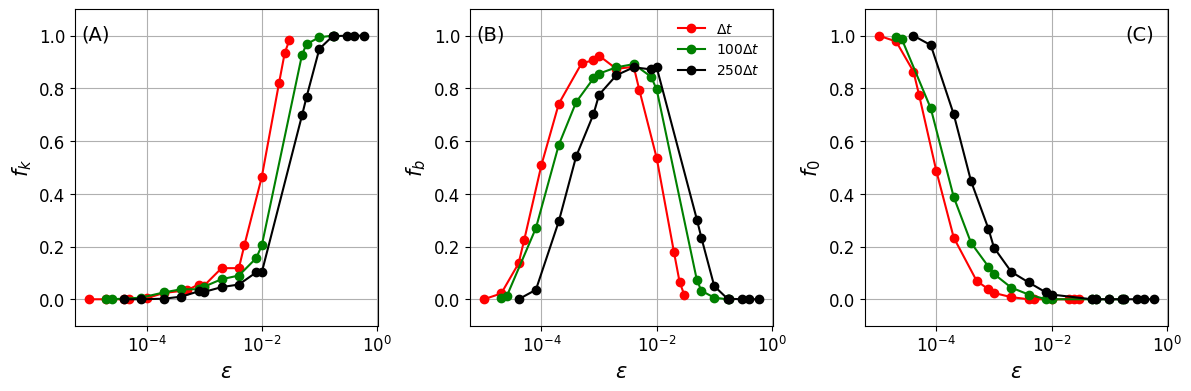}
     \caption{Occurrence frequencies of getting at least one kink-type object $f_k$, panel (A), breathers only $f_b$, panel (B), no excitations $f_0$, panel (C), versus the noise amplitude $\varepsilon$, for different values of the correlation time $\tau$. The different $\tau$ choices are $\Delta t$ (white-noise limit), $100\,\Delta t$, and $250\,\Delta t$, see panel (B) for the color scheme. Other parameters are: $\alpha = 0.2$, $A = 0.59$, $\omega = 0.6$, $L = 50$, $T = 500$ (simulation time), and $N = 500$ (number of trajectories).} 
     \label{fig2}
 \end{figure}

\begin{figure}[t!]
      \centering
      \includegraphics[width=1\linewidth]{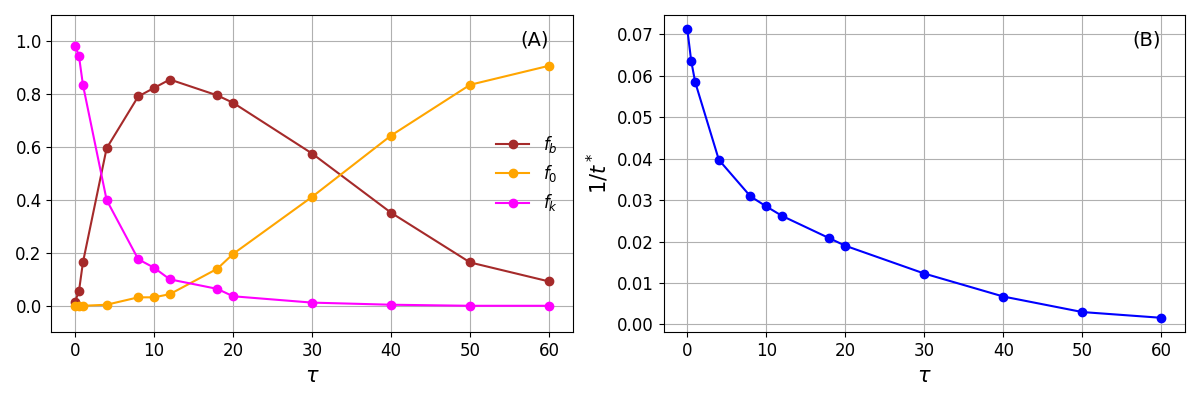}
      \caption{In panel (A) the occurrence frequencies are reported versus the correlation time $\tau$, starting from $\tau = \Delta t = 0.01$, as follows: at least one kink $f_k$ in magenta, breathers-only $f_b$ in brown, and no excitations $f_0$ in orange. In panel (B) the inverse hitting time is plotted as a function of $\tau$. Parameters: $\alpha = 0.2$, $A = 0.59$, $\omega = 0.6$, $\varepsilon = 0.04$, $L = 50$, $T = 500$, $N = 250$.}
      \label{fig3}
  \end{figure}

Let us first examine how the three occurrence frequencies behave, versus the noise amplitude $\varepsilon$, for different values of $\tau$. Starting from the white-noise limit, $\tau = \Delta t$, which fully recovers the results in Refs.~\cite{De_Santis_2023,De_Santis_2024_STAT}, we observe that for small $\varepsilon$ the frequency of kink-generation ($f_k$) and that of no-excitation generation ($f_0$) are, respectively, zero and one, see the red data points in Fig~\ref{fig2}(A) and (C). For large $\varepsilon$, the role of $f_k$ and $f_0$ is instead interchanged. Interestingly, $f_b$ shows a nonmonotonic behavior, with a clear peak identifying the optimal set of noise amplitudes for breather-related purposes, see the red curve in Fig~\ref{fig2}(B). That is, an intermediate $\varepsilon$ range exists where the most likely event is by far breather-only formation. The previous behaviors are well-understood: if the noise strength is too weak, creating localized excitations on top of the roughly uniform (oscillating) background is very unlikely, whereas, if it is too strong, all breathers are broken up into kink-antikink pairs~\footnote{Recall indeed that, according to the general SG theory, breathers represent kink-antikink bound states, see Ref.~\cite{Scott_2003,Dauxois_2006}}. Then, a sweet spot for which $\varepsilon$ is large enough for the excitation of breathers, but not for their destruction, allows for the peak in the red curve in Fig~\ref{fig2}(B) to develop. Beyond the white-noise limit, i.e., by increasing $\tau$, we observe that the previous picture still qualitatively holds, see the green ($\tau=100\,\Delta t$) and black ($\tau=250\,\Delta t$) curves in Fig~\ref{fig2}, thereby showing the robustness of the nonmonotonic $f_b$ behavior against temporal correlations in the noise source. Furthermore, Fig~\ref{fig2} clearly shows that the parameter $\tau$ shifts the occurrence frequency curves, thus dictating the range of $\varepsilon$ values where, for instance, the breather modes are most likely to be excited. The latter phenomenon, which can potentially be very useful in practice since it provides an additional control knob for breather generation, is intuitively explained by the above argument on the $\tau$-dependent noise amplitude rescaling.

It is interesting to note that, within the current framework, it is possible to leverage the effect of the noise's time correlations to enhance the likelihood of breather formation in a situation where, say, the value of $\varepsilon$ would only allow for kink-type excitations in the corresponding white-noise case. This can be seen in the $f_k(\tau)$, $f_b(\tau)$, and $f_0(\tau)$ curves in Fig.~\ref{fig3}(A), where we choose $\varepsilon = 0.04$, a value high enough to essentially result in $f_k\approx1$ at $\tau=\Delta t$ (white-noise limit, see the magenta curve). Notably, the breather only occurrence frequency [denoted in brown in Fig.~\ref{fig3}(A)], which starts from zero at $\tau=\Delta t$, is led to a maximum of $\sim 0.9$ through $\tau$'s increase, and eventually approaches again zero for very large values of $\tau$, since the latter are expected to effectively freeze the noisy dynamics. This phenomenon results in the concurrent increase of $f_0$, see the orange curve, whereas $f_k$ shows a decreasing trend, see the magenta curve. Overall, $f_b$ is found to be a nonmonotonic function of the correlation time. The freezing phenomenon can also be clearly appreciated by looking at the hitting time $t^\star$, see Fig.~\ref{fig3}(B). In particular, the latter figure, obtained for the same parameter set as Fig.~\ref{fig3}(A), depicts the average inverse hitting time, which is characterized by a monotonically decreasing behavior versus $\tau$. This quantity directly shows that progressively longer times are needed for the noise to trigger nonlinear modes as the correlation time increases, confirming the remarks made at the beginning of this section.

\subsection{Effects of spatially correlated noise}
\label{space}

\begin{figure}[t!]
    \centering
    \includegraphics[width=1\linewidth]{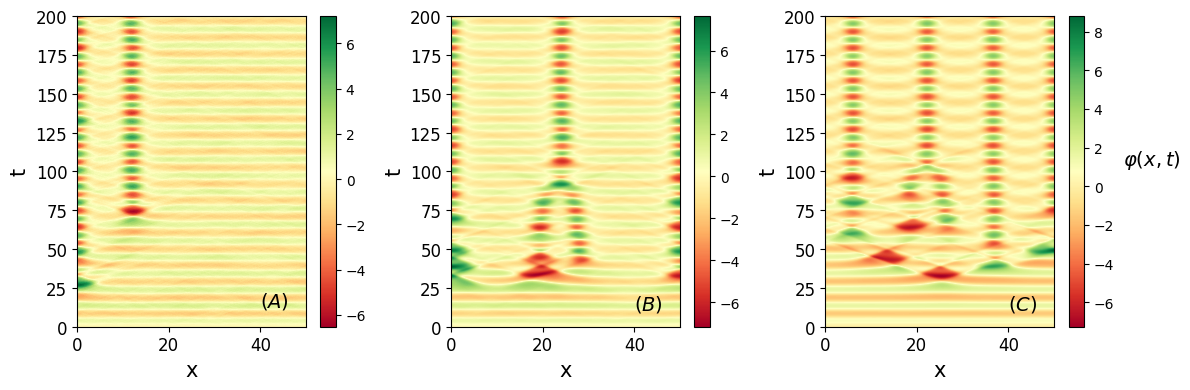}
    \caption{Contour plots of the field $\varphi(x, t)$, for different choices of the correlation length $\lambda$, exemplifying the robustness of the breather emergence phenomenon to spatial correlations, as well as the $\lambda$-induced collective excitation mechanism. We take $\lambda = \Delta x$ (i.e., white-noise limit) in panel (A), $\lambda = 1$ in panel (B), and $\lambda = 10$ in panel (C). The remaining simulation parameters are $\alpha = 0.2$, $A = 0.59$, $\omega = 0.6$, $\varepsilon = 0.004$. }
    \label{fig4}
\end{figure}

In this section, we investigate the effects of spatially correlated noise with correlation length $\lambda$, whereas delta-correlation in time is henceforth restored. We first illustrate, in Fig.~\ref{fig4}, the typical spatio-temporal behavior of the field $\varphi\of{x, t}$, for different values of $\lambda$. Specifically, in panel (A) we consider $\lambda = \Delta x$, corresponding to the white-noise limit in our discretized realm, in panel (B) $\lambda = 1$, and in panel (C) $\lambda = 10$. Similarly to the previous section, all other parameters are chosen such that the results of Ref.~\cite{De_Santis_2023,De_Santis_2024_STAT} are gradually approached as $\lambda \rightarrow 0$.

From these panels one immediately sees that the formation of breathers is also possible when spatial correlations in the noise are accounted for. Furthermore, the excited breathers remain stable both in amplitude and position over very long times. Notably, as compared to spatially uncorrelated scenarios, which seem to be characterized by generation events occurring in an essentially isolated fashion [cfr., e.g., Fig.~\ref{fig1} and Fig.~\ref{fig4}(A)], the early dynamical stages can display a peculiar collective behavior, intuitively when $\lambda$ exceeds the SG (anti)kink width of $\sim 1$ [cfr. Fig.~\ref{fig4}(C)], with similar solitonic structures emerging throughout. Thus, as opposed to both the delta- and self-correlated in time cases, spatially correlated sources are found to favor the formation of multiple excitations: as shown in the example in Fig.~\ref{fig4}(C), once the first solitons appear, the spatial correlations facilitate the spreading of such profiles, as witnessed by the activation of several nonlinear modes within the system. Remarkably, these convoluted excitation patterns can collapse into a bunch of isolated breathers persisting in time and space, as we observe in Fig.~\ref{fig4}(C).

\begin{figure}[t!]
    \centering
    \includegraphics[width=1\linewidth]{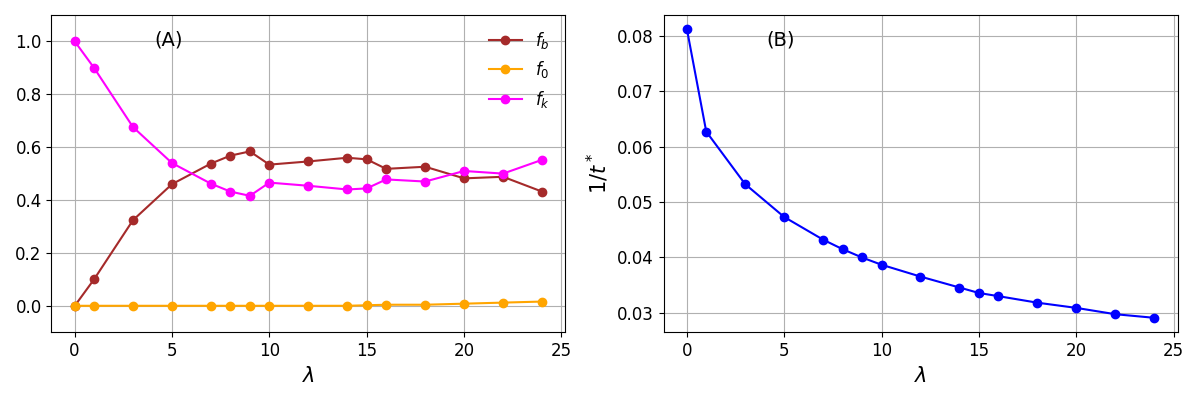}
    \caption{Panel (A): occurrence frequencies versus the correlation length $\lambda$, starting from $\lambda = \Delta x = 0.01$ (at least one kink $f_k$ in magenta, breathers-only $f_b$ in brown, and no excitations $f_0$ in orange). Panel (B): inverse hitting time as a function of $\lambda$. Parameters: $\alpha = 0.2$, $A = 0.59$, $\omega = 0.6$, $\varepsilon = 0.04$, $L = 50$, $T = 500$, $N = 250$.}
    \label{fig5}
\end{figure}

To further characterize the system's behavior in the spatially correlated case, we again turn to the statistical quantities introduced in the previous section, i.e., the occurrence frequencies $f_k$, $f_b$, and $f_0$ and the hitting time $t^\star$. In particular, we concentrate on an $\varepsilon$ value large enough, in a pure white-noise scenario, to (almost) always yield at least one kink-type excitation, i.e., for which $f_{k}\approx1$ and $f_{b,0}\approx0$, see also the red curves in Fig.~\ref{fig2}. Panel (A) of Fig.~\ref{fig5} displays the $f_k$, $f_b$, and $f_0$ occurrence frequencies as a function of the correlation length $\lambda$, see~\ref{app-b} for a discussion on the $\lambda$ range chosen for this investigation. As expected, for $\lambda\approx0$, the values $f_{k}\approx1$ [see the magenta curve in Fig.~\ref{fig5}(A)] and $f_b\approx f_0\approx0$ [see the brown and orange curves in Fig.~\ref{fig5}(A)] are observed. For higher $\lambda$, interestingly, we notice a decrease in the kink occurrence $f_k$ in favor of the breather one $f_b$, leading to the peak value $f_b\approx0.6$. The $f_b$ curve eventually starts going down, whereas $f_0$ is essentially equal to zero---up to $\lambda \approx L/2$, where it shows a small, but appreciable, increasing trend. The presence of spatial correlations in the noise can thus be exploited as well to obtain breather-only formation with a higher chance, as compared to the thermal case. On the other hand, Fig.~\ref{fig5}(B) illustrates the inverse hitting time $1/t^\star$ versus $\lambda$, which is characterized by a monotonically decreasing behavior.

Both the plots in Fig.~\ref{fig5} contain valuable information about the dynamical behavior of the system. Increasing the $\lambda$ parameter results in a rescaled effective noise amplitude, similarly to the time-correlated case, which in turn explains the observation of a dynamics' slowdown in Fig.~\ref{fig5}(B). However, differently from the previous section, where the freezing phenomenon is found in correspondence of a decrease in both the kink and breather occurrence, now the spatial correlations seem to lead to an effective exchange between $f_k$ and $f_b$, see Fig.~\ref{fig5}(A). We can intuitively understand the persistence in the $f_{k,b}$ occurrence frequencies versus $\lambda$ as follows: although the dynamics is progressively slowing down, larger $\lambda$ values imply easier-to-spread fluctuations throughout the system and thus trigger collective behavior, such that the effectiveness of the noisy generation events can still be very high. We eventually expect $f_0$ to increase, as $\lambda \rightarrow \infty$, since in this limit we recover a spatially uniform system. In this regard, a hint is provided by the slightly increasing behavior of the $f_0$ (orange) curve as we reach $\lambda=L/2$.

\section{Conclusions}
\label{end}

This work concerns the effects of both temporally and spatially correlated noise sources on the emergence of SG breather modes robust against dissipation, in the presence of spatially uniform ac driving. We find both the correlation time $\tau$ and the correlation length $\lambda$ to influence the likelihood of observing breathers, as well as the timescale for the excitation of these nonlinear modes. Specifically, in the $\tau\to0$ (white noise) and the $\lambda\to0$ (spatially uncorrelated noise) limits, we recover the results from Refs.~\cite{De_Santis_2023,De_Santis_2024_STAT}; on the other hand, for both large $\tau$ and $\lambda$ we observe a progressive dynamical slowdown, that is, noise-induced nonlinear modes take longer times to arise. We find, notably, that intermediate ranges of correlation times and correlation lengths exist such that breathers are observed much more frequently as compared to the corresponding white, spatially uncorrelated, noise case. In other words, the breather-only occurrence frequency shows nonmonotonicities as a function of both $\tau$ and $\lambda$. We also note interesting qualitative differences in the events leading to the formation of solitonic structures between the temporally and the spatially correlated scenarios. In the former case, indeed, localized modes are generated in an essentially isolated fashion, i.e., thanks to intense enough fluctuations of the field occurring over a relatively small portion of the spatial domain, whereas in the latter case (for large enough $\lambda$) we observe a tendency towards collective behaviors, with easy-to-spread fluctuations leading to a cascade of solitons emerging throughout the system. Nevertheless, these rather convoluted transients can still relax into a number of isolated breathers, robust over very long times.

Given that noisy and ac-driven LJJs have been recently proposed as ideal candidates for the experimental observation of breathers, via both destructive~\cite{De_Santis_2022,De_Santis_2022_CNSNS,De_Santis_2023} and non-destructive~\cite{De_Santis_2024_HEAT} approaches, the above results are expected to be useful in view of the actual implementation of such detection protocols, thereby solving a long-standing problem in the Josephson community~\cite{Parmentier_1993,Gulevich_2012,Monaco_2019}. We have, in fact, uncovered additional knobs for the controlled excitation of these elusive nonlinear modes by exploiting the correlation properties of the noise source. As a further step, in the future it might be interesting to concentrate on non-Gaussian fluctuations as well, such as Lévy-distributed noise sources~\cite{Chechkin_2007,Afek_2023,Baule_2023}, and establish their effects on the current framework. In particular, Lévy sources are known to lead to rich dynamical phenomena in SG-type landscapes~\cite{Guarcello_2013,Guarcello_2020,Guarcello_2021}, and some questions naturally arise: for example, could the celebrated Lévy flights lead to more effective breather generation? Another aspect worth of further exploration is that of the collective solitonic behaviors induced by the spatial correlations in the noise. We expect these phenomena to be of general interest from the perspective of statistical physics in the SG domain~\cite{Fedorov_2007,Fedorov_2008,Pankratov_2008,Aug09,Pankratov_2012,Guarcello_2015_1}, even in Kibble-Zurek-like scenarios~\cite{Gordeeva2010}, in addition to the specific breather generation framework considered here. In this regard, the widespread interest of the scientific community in the physics of solitons~\cite{Belinski2001,Scott_2006,Purwins2010,Suchkov2016,Bykov_2016,Song2019,Wang2020,Yakushevich_2021,Malomed_2022,Malomed2024} could drive investigations into the peculiar features of solitonic structures' emergence and dynamics, particularly in the presence of spatially correlated noise sources, extending even beyond the SG model.

\section*{Acknowledgements}
The authors acknowledge the support of the Italian Ministry of University and Research (MUR). A.C. also acknowledges support from European Union – Next Generation EU through Project Eurostart 2022 (MUR D.M. 737/2021) and through Project PRIN 2022-PNRR no. P202253RLY.

\appendix

\section{Temporally correlated noise}
\label{app-a}
In this appendix, we will focus on a noise source displaying nonzero temporal correlations, but delta-correlated in space, i.e., with autocorrelation function given by
\begin{equation}
     \mn{\eta\of{x,t}\eta\of{x^\prime, t^\prime}} = \frac{\varepsilon}{\tau} \delta\of{x - x^\prime} e^{-\frac{|t-t^\prime|}{\tau}}, \label{A1}
\end{equation}
where $\varepsilon$ is the noise amplitude and $\tau$ is the correlation time. In short, Eq.~\eqref{A1} implies that the outcomes $\eta(x, t)$ and $\eta(x, t^\prime)$ are not independent of each other, and the overlap between the two exponentially decays with the time interval $|t-t^\prime|$ over the correlation time $\tau$. By contrast, at each time $t$, no correlation is displayed between noise outcomes at different locations $x$ and $x^\prime$, regardless of how short the distance $|x-x^\prime|$ is. In other words, the memory of the noise's history is lost, at any location $x$, over the characteristic time $\tau$. Moreover, from Eq.~\eqref{A1} we can infer the time correlation to introduce a rescaled noise amplitude, with the peculiarity that, as $\tau \rightarrow 0$, the white noise correlation properties are regained.

To accurately simulate the noise in Eq.~\eqref{A1} and its statistical properties, we analyze an Ornstein–Uhlenbeck process, similarly to Ref.~\cite{Garcia_2012}, which obeys
\begin{equation}
    \partial_{t}\eta\of{x, t} = -\frac{1}{\tau}\eta\of{x, t} + \frac{1}{\tau}\xi\of{x, t},\label{4}
\end{equation}
where $\xi\of{x,t}$ is a Gaussian noise source with zero mean and the following correlation function
\begin{equation}
    \mn{\xi\of{x,t}\xi\of{x^\prime, t^\prime}} = 2\varepsilon \delta\of{x  - x^\prime}\delta\of{t - t^\prime}.\label{5}
\end{equation}
It can be rigorously shown that the stochastic variable $\eta$ in Eq.~\eqref{4} follows the same statistics as that in Eq.~\eqref{A1}. Thus, by numerically integrating the Ornstein–Uhlenbeck process, we can generate the desired fluctuations' pattern. To this end, Eq.~\eqref{4} can by handled via a first-order Euler method, resulting in the discretized expression
\begin{equation}
    \eta_{i,j+1} = \of{1- \frac{\Delta t}{\tau}}\eta_{i,j} + \frac{\Delta t}{\tau}\frac{\sqrt{2\varepsilon}}{\sqrt{\Delta x \Delta t}} N_{i,j},
\end{equation}
where $i$ and $j$ are the indices relative to the spatio-temporal grid, characterized by the discretization steps $ \Delta x = \Delta t = 0.01$ for all of our simulations, and $N_{i,j}$ are independent normal random variables with zero mean and unit variance.
\begin{figure}[t!]
    \centering
    \includegraphics[width=1\linewidth]{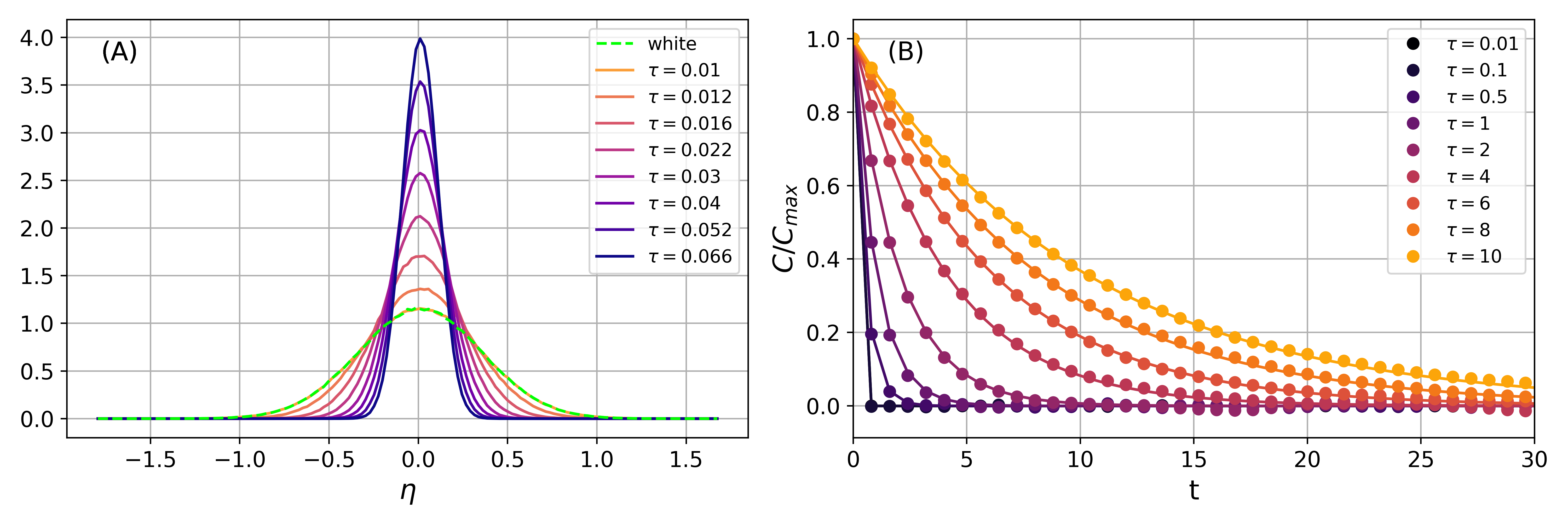}
    \caption{Panel (A) shows the histograms computed for the time-correlated noise sources, where distinct colors represent different values of the correlation time, as well as that for the corresponding white noise case, see the green curve. The histograms are normalized such that the area under each curve is the same. Panel (B) displays the correlation functions versus $t$ at different $\tau$, with dots indicating the numerical outcomes and solid lines representing the analytical predictions. The correlation functions are only plotted for a suitable time range (the simulation time is $T = 500$, the same as in the main text), but they can all be observed to approeach zero for large enough times. In both panels, the amplitude $\varepsilon = 0.06$ is taken. }
    \label{fig8}
\end{figure}

To quantitatively illustrate how $\tau$'s presence can alter the effective noise amplitude, Fig.~\ref{fig8}(A) displays the histograms computed for time-correlated noise sources (distinct colors represent different values of the correlation time), for the fixed value $\varepsilon = 0.06$, and compares them to the corresponding white noise case (in green). It is evident that the noise distribution remains Gaussian in all cases, but with a variance which decreases with $\tau$. The white-noise variance is recovered for $\tau = \Delta t$, i.e., the closest we can get to the time-uncorrelated limit in the discretized framework. Moreover, in Fig.~\ref{fig8}(B), we test the reliability of the above numerical scheme for time-correlated noise generation by comparing the simulated correlation functions (see the dots in the plot) against the corresponding analytical predictions (see the solid lines in the plot), for different values of the correlation time. We observe a very good agreement throughout the explored parameter range. Finally, we remark that our simulations, discussed in the main text, all involve values of $\tau$ much smaller than the typical simulation time, $T=500$, such that the effects of time correlations on the system's dynamics can be addressed in a meaningful way.

\section{Spatially correlated noise}
\label{app-b}
Here we will concentrate on a noise source with nonzero spatial correlations, but uncorrelated in time, characterized by the autocorrelation function
 \begin{equation}
    \mn{\eta\of{x,t}\eta\of{x^\prime, t^\prime}} = \frac{\varepsilon}{\lambda}e^{-\frac{|x-x^\prime|}{\lambda}} \delta\of{t - t^\prime},\label{b1}
 \end{equation}
where $\varepsilon$ is the noise amplitude and $\lambda$ is the correlation length. In short, Eq.~\eqref{b1} implies that the outcomes $\eta(x, t)$ and $\eta(x^\prime, t)$ are not independent of each other, and the overlap between the two exponentially decays with the distance $|x-x^\prime|$ over the characteristic scale $\lambda$. By contrast, at each point $x$, no correlation is displayed between noise outcomes at different times $t$ and $t^\prime$, regardless of how short the interval $|t-t^\prime|$ is. Additionally, from Eq.~\eqref{b1} we see that the presence of spatial correlations results in a rescaled effective noise amplitude, and the white noise correlation properties are recovered as $\lambda \rightarrow 0$.

We now turn to a technical description on how to numerically simulate a noise source delta-correlated in time, with nonzero spatial correlations, according to the definition in Eq.~\eqref{b1}. Following Ref.~\cite{Garcia_2012}, we employ an algorithm that allows for the generation of spatially correlated noise with the following correlation function 
\begin{equation}
    \mn{\eta\of{x, t} \eta\of{x^\prime, t^\prime}} = 2C\of{\frac{\left\vert x - x^\prime \right\vert}{\lambda}}\delta\of{t - t^\prime},\label{6}
\end{equation}
where $\varepsilon$ is the noise amplitude, $\lambda$ is the correlation length, and $C(x)$ is an arbitrarily shaped correlation function [for our numerical runs, an exponential profile is taken, see Eq.~\eqref{b1}]. This particular type of noise can be handled in practice by working in Fourier space. The Fourier transform of $\eta(x, t)$ can be written as
\begin{equation}
 \eta\of{k, t} = \sqrt{C\of{k}}\alpha\of{k, t},\label{7}  
\end{equation}
where $C(k)$ is the Fourier transform of $C(x)$ and $\alpha(k, t)$ are random Gaussian numbers with zero mean and correlation function given by
\begin{equation}
      \mn{\alpha\of{k, t} \alpha\of{k^\prime, t^\prime}} = 2\delta\of{k + k^\prime}\delta\of{t - t^\prime}.\label{8}
\end{equation}
The noise in Fourier space can be simulated in two different ways. The first, though not very efficient, is to generate white noise in real space and then compute its Fourier transform. Another possibility, which is pursued here, is to generate the noise directly in Fourier space. To do so, one can straightforwardly show that the Fourier transform of white real-space noise inherits the following correlation property
\begin{equation}
    \mn{\alpha\of{k}\alpha\of{k^\prime}} = \delta\of{k + k^\prime},\label{9}
\end{equation}
\sloppy and that both the relations Re$\left\lbrace\alpha(k)\right\rbrace$=Re$\left\lbrace\alpha(-k)\right\rbrace$ and Im$\left\lbrace\alpha(k)\right\rbrace$=--Im$\left\lbrace\alpha(-k)\right\rbrace$ hold. To generate the noise signal in Eq.~\eqref{9}, the field $\alpha(k)$ can be expressed in terms of its real and imaginary parts $\alpha(k) = a(k) + ib(k)$. Here, $a$ and $b$ are two random Gaussian variables with zero mean and variances $\langle a^2 \rangle = \langle b^2 \rangle = \frac{1}{2}$ for every point in the Brillouin zone, except for the edge and the middle point, where $\langle a^2 \rangle = 1$ and $\langle b^2 \rangle = 0$.

\begin{figure}[t!]
    \centering
    \includegraphics[width=1\linewidth]{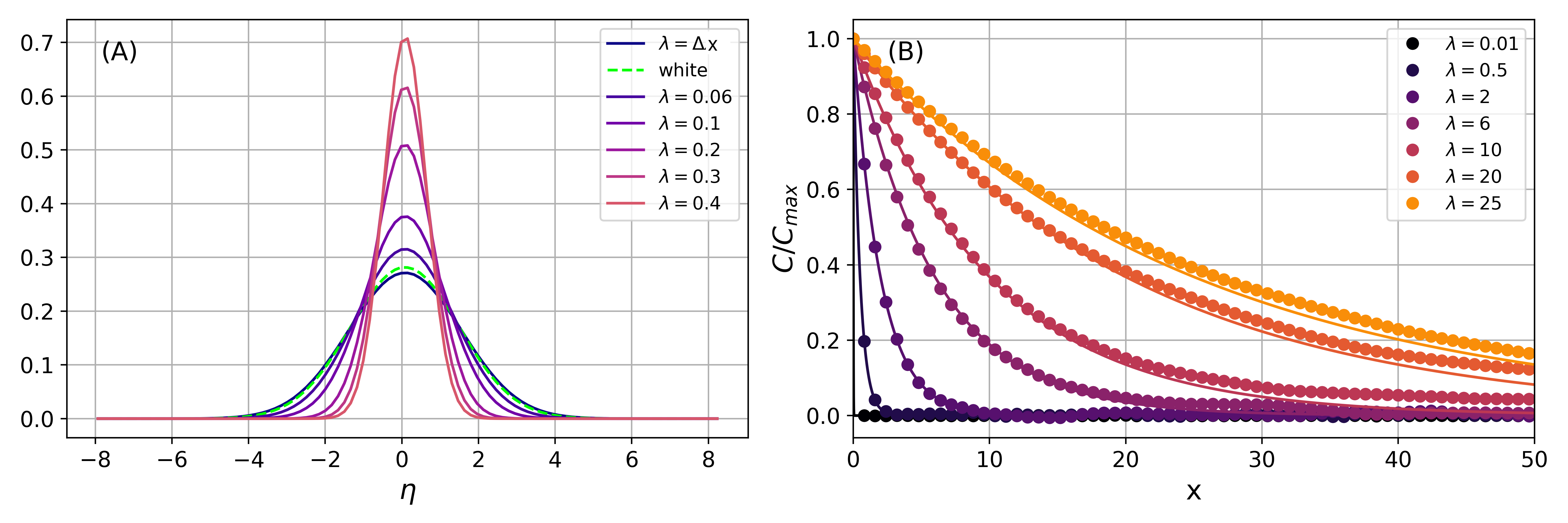}
    \caption{Panel (A) shows the histograms computed for the spatially correlated noise sources, where distinct colors represent different values of the correlation length, as well as that for the corresponding white noise case, see the green curve. The histograms are normalized such that the area under each curve is the same. Panel (B) displays the correlation functions versus $x$ at different $\lambda$, with dots indicating the numerical outcomes and solid lines representing the analytical predictions. In both panels, the amplitude $\varepsilon = 0.001$ is taken.}
    \label{fig10}
\end{figure}
To address the effectiveness of our noise generation scheme, we perform a statistical analysis similar to that presented in~\ref{app-a}. We first examine the Gaussian nature of the noise and its rescaling properties. Figure~\ref{fig10}(A) illustrates the histograms computed for spatially correlated noise sources (distinct colors represent different values of the correlation length), for the fixed value $\varepsilon = 0.001$, and compares them to the corresponding white noise case (in green). It is evident that the noise distribution remains Gaussian regardless of $\lambda$, but progressively smaller variances are obtained as the correlation length grows. We recover the white noise variance for $\lambda = \Delta x$, i.e., the closest the discretized framework can get to the length-uncorrelated limit.

Finally, in Fig.~\ref{fig10}(B) we compare the correlation functions obtained from the numerical trajectories with their exact analytical counterparts, given by Eq.~\eqref{b1}, for different choices of the correlation length $\lambda$. We observe a good agreement for both small and intermediate values of $\lambda$, whereas deviations between the analytical prediction and the numerical results start to occur when $\lambda$ is increased to values comparable to the length $L$ of the system due to finite-size effects. In light of this, we have chosen $\lambda = 25$ as the maximum correlation length for our computational experiments.

\end{document}